\documentclass[%
 reprint,
superscriptaddress,
nofootinbib,
 amsmath,amssymb,
 aps,
floatfix
]{revtex4-2}

\usepackage[breaklinks,colorlinks]{hyperref}
\usepackage{xcolor}
\usepackage{soul}
\usepackage{graphicx}
\usepackage{dcolumn}
\usepackage{bm}
\usepackage{tikz}
\usetikzlibrary{quantikz2}
\newcommand\numberthis{\addtocounter{equation}{1}\tag{\theequation}}

\begin{document}


\title{Variance Minimisation of the Lipkin-Meshkov-Glick Model\\ on a Quantum Computer}

\author{I. Hobday}
\affiliation{School of Mathematics and Physics, University of Surrey,\\ Guildford, Surrey, GU2 7XH, United Kingdom}
\author{P. D. Stevenson}
\affiliation{School of Mathematics and Physics, University of Surrey,\\ Guildford, Surrey, GU2 7XH, United Kingdom}
\affiliation{AWE, Aldermaston, Berkshire, RG7 4PR, United Kingdom }
\author{J. Benstead}
\affiliation{AWE, Aldermaston, Berkshire, RG7 4PR, United Kingdom }
\affiliation{School of Mathematics and Physics, University of Surrey,\\ Guildford, Surrey, GU2 7XH, United Kingdom}

\date{\today}

\begin{abstract}
{Quantum computing can potentially provide advantages for specific computational tasks. The simulation of fermionic systems is one such task that lends itself well to quantum computation, with applications in nuclear physics and electronic systems. Here we present work in which we use a variance minimisation method to find the full spectrum of energy eigenvalues of the Lipkin-Meshkov-Glick model; an exactly-solvable nuclear shell model-type system.

We perform these calculations using both quantum simulators and real quantum hardware accessed via IBM cloud-based quantum computers. Using these IBM quantum computers we are able to obtain all eigenvalues for the cases of three and seven fermions (nucleons) in the Lipkin-Meshkov-Glick model.} 
\end{abstract}

\maketitle


\section{\label{sec:level1}Introduction}

The use of quantum algorithms to perform calculations in nuclear physics is a rapidly developing field \cite{visnak_quantum_2015,visnak_quantum_2017,dumitrescu_cloud_2018,roggero_preparation_2020,roggero_quantum_2020,lacroix_symmetry-assisted_2020,siwach_filtering_2021,baroni_nuclear_2021,robin_entanglement_2021,zhang_selected_2021,robbins_benchmarking_2021,perez-fernandez_digital_2022,cervia_lipkin_2021,kiss_quantum_2022,romero_solving_2022,chikaoka_quantum_2022,hlatshwayo_simulating_2022,ruiz_guzman_accessing_2022,hobday,PhysRevC.105.064308,li2023quantum}.  The exponential scaling of Hilbert space with the number of quantum bits (qubits) and the ability of multiple qubits to exhibit highly entangled wave functions give quantum computers the potential to have a great impact in simulating many-body quantum systems. In particular, one can expect a \textit{quantum advantage} where quantum computation outperforms classical computation as the size of the system under study becomes sufficiently large.  

Richard Feynman first proposed using quantum computers to study quantum systems in the 1980s \cite{feynman_simulating_1982}. Since then, various algorithms have been developed that are able to perform calculations on many-body quantum systems, such as Quantum Phase Estimation (QPE) \cite{OBrien_2019,zhou2013calculating,du2021ab}, and Quantum Imaginary Time Evolution \cite{Motta2020,PhysRevA.105.022440,PhysRevA.106.062435}. However, these algorithms are often too complex to be fully implemented on current quantum computers. 

Current quantum computers are said to be in the Noisy, Intermediate-Scale Quantum (NISQ) era. This is due to the low numbers of qubits, and large amounts of noise and error present within the devices. Current devices also have low coherence times, restricting the amount of time a qubit can maintain its state. This limits the length of the quantum circuit, or circuit depth, that quantum computers can use and still produce meaningful results \cite{Preskill_2018}. 

\subsection{\label{sec:variational_algorithms}Variational Algorithms}
Variational quantum algorithms have established themselves due to the relative ease of running them on NISQ-era quantum computers \cite{Peruzzo2014,McClean_2016,tilly2022variational}. These are hybrid algorithms, meaning they use both  quantum and classical processes in tandem, reducing the amount of computational work needed to be performed on the quantum computer. Working in conjunction with a classical computer allows for reduced circuit depths and improved error rates by reducing the number of gates in the quantum circuit. This reduction in error allows for meaningful results to be obtained on current quantum hardware.

The Variational Quantum Eigensolver (VQE) algorithm uses the variational principle of quantum mechanics to approximate the ground state of some Hamiltonian. In the work presented here, we apply a version of the VQE which targets any eigenvalue of a quantum system by minimizing the variance of the Hamiltonian rather than the expectation value. Application of the method is made to the simplified nuclear model due to Lipkin, Meshkov, and Glick, which we summarize in the next section.  The presentation proceeds with a description of our implementation of the algorithm, followed by results on simulated and real quantum computers.

\subsection{\label{subsec:lipkin}Lipkin-Meshkov-Glick Model}

The Lipkin-Meshkov-Glick (LMG) model is an exactly solvable nuclear model introduced in the 1960s \cite{LIPKIN1965188}. It is a simple shell model consisting of two levels, separated by an energy $\epsilon$ with a permutation-symmetric potential.  The two levels are each $N$-fold degenerate and a standard treatment (which we adopt here) considers $N$ fermions in the system. The Hamiltonian for the model is given as

\begin{eqnarray}
    H =&& \frac{1}{2}\epsilon \sum_{p\sigma}\sigma a_{p\sigma}^\dagger a_{p\sigma}+\frac{1}{2}V\sum_{pp'\sigma} a_{p,\sigma}^\dagger a_{p',\sigma}^\dagger a_{p',-\sigma}a_{p,-\sigma}\nonumber\\
    &&+ \frac{1}{2}W \sum_{pp'\sigma}a_{p,\sigma}^\dagger a_{p',-\sigma}^\dagger a_{p',\sigma} a_{p,-\sigma} \label{eq:LipkinOri}.
\end{eqnarray}

Here the labels $p$ and $p'$ run over the $N$ degenerate states in each level, while $\sigma=\pm 1$ labels the level.  The strength $V$ controls pair (de-)excitations between the two levels, while the $W$ term scatters one particle up and another down. 
The model can be simplified by writing the Hamiltonian in the \textit{quasi-spin} basis, as

\begin{equation}
    H = \epsilon J_z + \frac{1}{2}V(J_+^2 + J_-^2) + \frac{1}{2}W(J_+J_- + J_-J_+)\label{eq:Lipquasi},
\end{equation}

\noindent where $J_z$ and $J_\pm$ are given by 

\begin{equation}
    J_z = \frac{1}{2}\sum_{p\sigma} \sigma a_{p\sigma}^\dagger a_{p\sigma}
\end{equation}
and
\begin{equation}
    J_\pm = \sum_p a_{p\pm1}^\dagger a_{p\mp 1} 
\end{equation}

\noindent respectively.

These quasi-spin operators satisfy angular momentum commutation relations.  The total quasi-spin operator $J^2=\frac{1}{2}(J_+J_-+J_-J_+)+J_z^2$ commutes with the Hamiltonian, so that each $J$ value may be considered separately.  This symmetry allows one to reduce the maximum Hamiltonian matrix dimension from $2^N$ to $N+1$. Further simplification is possible since the Hamiltonian also exhibits good ``parity'' \cite{agassi_validity_1966}, which splits each matrix of a given $J$ into two block-diagonal parts.

The LMG model is applicable as a test-platform for quantum computing algorithms applied to nuclear physics due to its exactly solvable nature, and the ability to scale the model indefinitely by increasing the number of particles in the system.  The model has already been studied using real and simulated quantum computers, particularly for calculation of its ground state \cite{cervia_lipkin_2021,chikaoka_quantum_2022}, but also for excited states using a quantum-assisted algorithm to prepare a generalised eigenvalue problem for solution on a classical computer for the excited state spectrum \cite{hlatshwayo_simulating_2022}. For our work, we use an LMG model Hamiltonian with the parameters $\frac{V}{\epsilon}=0.5$, $\frac{W}{\epsilon}=0$, making the common choice to ignore the $W$ term which does not affect ground state correlations.

\section{\label{sec:encoding}Encoding Methodology}

The LMG-model Hamiltonian must be encoded in such a way as to allow it to be processed by a quantum computer. For circuit-based quantum computers, this form is conveniently expressed as linear combinations of Pauli spin matrices. 
Often a representation of a Hamiltonian in second-quantized form, such as Eq. (\ref{eq:LipkinOri}), leads to a mapping from creation and annihilation operators to Pauli strings, keeping the second quantized representation where solutions of any particle number can be encoded.  Such encodings as Jordan-Wigner \cite{Jordan1928} or Bravyi-Kitaev \cite{BRAVYI2002210,Bravyi} fall in this category.  Having made use of the quasispin representation, and specialising to the standard Lipkin Model in which the number of particles is fixed at $N$ - the same as the degeneracy of each of the two unperturbed levels - we make a more specialised and efficient encoding starting from a representation of the Hamiltonian in the quasispin basis \cite{cervia_lipkin_2021} in which only fixed-particle-number states are considered.  

For a given specific LMG Hamiltonian, $H_N$ with fixed $N$ we make a matrix representation in the quasispin basis and consider (as usual) the maximum quasispin value, that being the one containing the unperturbed ground state.  The matrix dimension is $N+1$, but owing to the parity symmetry, it can be written as two block diagonal parts which can be treated independently.  When $N$ is odd, the two submatrices are of equal dimension, $(N+1)/2$.  In the present work we consider $N=3,7$ so that $(N+1)/2$ is of the form $2^m$ and each matrix whose eigenvalues are sought has a power-of-two dimension.  For a Hamiltonian $H_N$ for which $2^m=(N+1)/2$ we can decompose in terms of Pauli matrices as

\begin{equation}
    H_N = \sum_{i=0}^{2^{2m}-1}\beta_i\bigotimes_{q=1}^m \sigma_{{i_4}[q]}
\end{equation}

\noindent
where $\beta_i$ is a coefficient and the tensor product $\bigotimes$ produces all Pauli strings from the set $\{\sigma_0,\sigma_1,\sigma_2,\sigma_3\}=\{I,X,Y,Z\}$ and the notation $\sigma_{{i_4}[q]}$ means the $q^{th}$ digit from the right of $i$ when represented in base 4.

For the given numerical Hamiltonians with particular values for the interaction strengths $V/\epsilon$ and $W/\epsilon$, the coefficients $\beta_i$ were found using a custom code \cite{pesce_h2zixy_2021}, whose functionality is also found in common quantum computing libraries (e.g. the \texttt{pauli\_decompose} method of PennyLane \cite{bergholm2022pennylane}).  Specific examples of the encodings are found in subsequent sections.





\section{\label{sec:excited}Finding Excited States}

In a typical Variational Quantum Algorithm (VQA), an ansatz is made for a wave function representing the ground state of the problem at hand.  The ansatz contains parameters which can be adjusted to optimise the wave function and bring it as close as possible to the true ground state.  Usually the expectation value of the Hamiltonian, expressed in terms of Pauli operators, is evaluated on the quantum computer, while a classical algorithm manages the parameter optimisation.

Here, we adapt the typical VQA to seek the minimum of the variance of the Hamiltonian,

\begin{equation} \label{eq:variance}
    \sigma^2 = \langle H^2 \rangle - \langle H \rangle^2.
\end{equation}

The variance is a positive semi-definite function which is zero when the wave function is an eigenfunction of the Hamiltonian.  Hence, its set of equally deep global minima will correspond to the set of eigenstates of the Hamiltonian, except possibly for accidental zeros of the variance which can be checked for. 

This method allows the use of the same quantum circuit ansatzes as when using the VQE to find the ground state of $H$ directly, but requires additional circuit measurements to be performed for the terms in $\langle H^2 \rangle$. This increases the time taken to perform calculations but does not introduce any additional circuit depth or variational parameters to the calculations compared with an energy-minimizing ground state VQE. Since no additional circuit complexity is introduced, using the variance to find excited states of the nuclear system is possible on NISQ-era quantum hardware for problems where the standard VQE is applicable.

In our implementation, we make use of the IBM qiskit environment to perform our VQE on real quantum hardware, as noted later where individual results are presented, using qiskit's classical COBYLA solver for the ansatz parameter optimisation.

\section{\label{sec:Simulation_Meas} \texorpdfstring{$N=3$}{N=3} Simulation and Measurement}
\subsection{\texorpdfstring{$N=3$}{N=3} Hamiltonian}
The Hamiltonian for the $N=3$ LMG model is split into its two parity submatrices labelled as $A$ and $B$.  For our chosen values of $V/\epsilon=0.5$ and $W/\epsilon=0$ the submatrices can be expressed as

\begin{equation}
    H_{N=3 A} = \begin{bmatrix}
        -1.5 & -0.866 \\
        -0.866 & 0.5
    \end{bmatrix} \label{eq:N=3A}
\end{equation}
and 
\begin{equation}
    H_{N=3 B} = \begin{bmatrix}
        -0.5 & -0.866 \\
        -0.866 & 1.5
    \end{bmatrix}. \label{eq:N=3B}
\end{equation}
When decomposed into the form of Pauli matrices, the Hamiltonians can be represented as

\begin{equation}\label{eq:H3APauli}
    H_{N=3A} = -0.5 - 1.0 Z_0 - 0.8660254 X_0 
\end{equation}
and 
\begin{equation}\label{eq:H3BPauli}
    H_{N=3B} = 0.5 - 1.0 Z_0 - 0.8660254 X_0 
\end{equation}
respectively, where we have kept more significant figures than given in the matrix representation.

We note that the two submatrices differ only by the signs of all the diagonal elements, while the off-diagonal elements are indetical.  This means that the eigenspectra of the two submatrices differ only by their sign.  This is a general result (for odd $N$) and one needs only compute the spectrum of one of the submatrices in order to know the spectrum for both, but we test the method in the N=3 case for both submatrices. 

For the evaluation of the variance, Eq. (\ref{eq:variance}), the square of the Hamiltonian operator is needed.  From the square of the matrix representations

\begin{equation}
    H^2_{N=3 A} = \begin{bmatrix}
        1.0 & 0.866 \\
        0.866 & 3.0
    \end{bmatrix} \label{eq:N=3Asq}
\end{equation}

and 

\begin{equation}
    H_{N=3 B}^2 = \begin{bmatrix}
        1.0 & -0.866 \\
        -0.866 & 3.0
    \end{bmatrix} \label{eq:N=3Bsq}
\end{equation}

their Pauli representations are found as

\begin{equation}
    H^2_{N=3A} = 2.0 + 0.8660254 X_0 + 1.0 Z_0
    \end{equation}
and
    \begin{equation}
H^2_{N=3B}=2.0 - 0.8660254 X_0 - 1.0 Z_0
\end{equation}
\subsection{\label{sec:ansatz}Circuit Ansatz}

\begin{figure}[tbh]
    \centering
    \begin{quantikz}
        \\
        \lstick{\ket{0}} & \qw & \gate{R_y(\theta)} & \qw  & \qw \\
		\end{quantikz}
    \caption{\label{fig:1q_circuit_ansatz}Single-qubit circuit ansatz.}        
\end{figure}
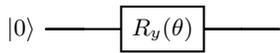

From the submatrices presented in Eqs. (\ref{eq:N=3A}) and (\ref{eq:N=3B}), or their Pauli form in Eqs. (\ref{eq:H3APauli}) and (\ref{eq:H3BPauli}), it is clear that the wavefunctions of these submatrices can be expressed using a single qubit ansatz. Thus we use a one-parameter circuit, shown in Fig. \ref{fig:1q_circuit_ansatz}, to cover the Hilbert space to find both eigenvalues associated with the respective submatrix being measured.


\subsection{Simulation}\label{sec:LMG3_simulation}



\begin{figure}[tbh]
    \centering
    \includegraphics[width=8.6cm,clip]{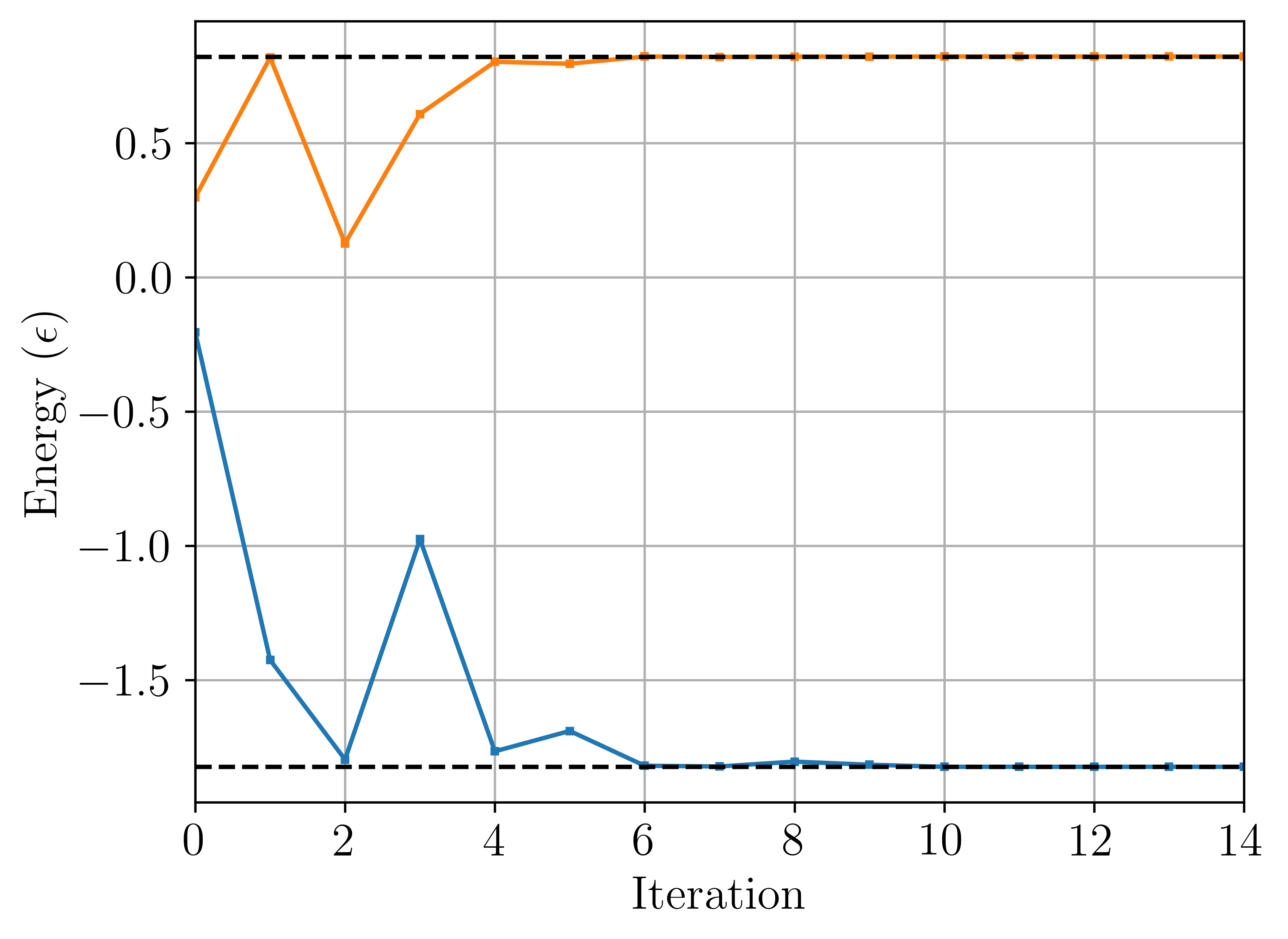}
    \caption{Iterative variance minimisation to converge on energy eigenstates (dashed lines) of the $H_{N=3A}$ sector of the $N=3$ LMG model with parameters $\frac{V}{\epsilon}=0.5$, $\frac{W}{\epsilon}=0$, using a simulator.}
    \label{fig:LMG3_sim}
\end{figure}

We perform variance minimisations, starting from random initial parameters, using noiseless simulations. The results of these simulations are plotted in Fig. \ref{fig:LMG3_sim} for the submatrix $H_{N=3A}$, which show the energy values converging to their respective exact eigenvalues. 

\subsection{\label{sec:Quantum Hardware Results}Quantum Hardware Results}

We perform the calculations of the $N=3$ LMG excited state spectrum using a parameter sweep, since this is feasible for a single parameter and will help visualise the results.  For this, we use IBM's cloud-based quantum computer, IMBQ\_manila. This is a 5-qubit quantum computer using superconducting circuits and a ``nearest neighbour" entanglement scheme (though for this $N=3$ case we use only a single qubit). We perform the calculations using 20,000 measurements per circuit (shots), and one circuit for each term in the Hamiltonians. We also introduce additional circuits to mitigate some of the measurement readout bias that is innate to the quantum computer. These additional circuits prepare a state that represents each possible output of the quantum computer, using Pauli X gates. For a single qubit, this method of mitigation adds two additional circuits, one measuring the state $|0\rangle$ and the other measuring the state $|1\rangle$, per iteration step of the VQE. This allows for up-to-date corrections to be applied to our counts \cite{Kandala2017}.  We also perform VQE variance minimisations, starting from random initial parameters using IBMQ\_manila as described in section \ref{sec:excited}.



\begin{figure}[tbh]
    \centering
    \includegraphics[width=8.6cm,clip]{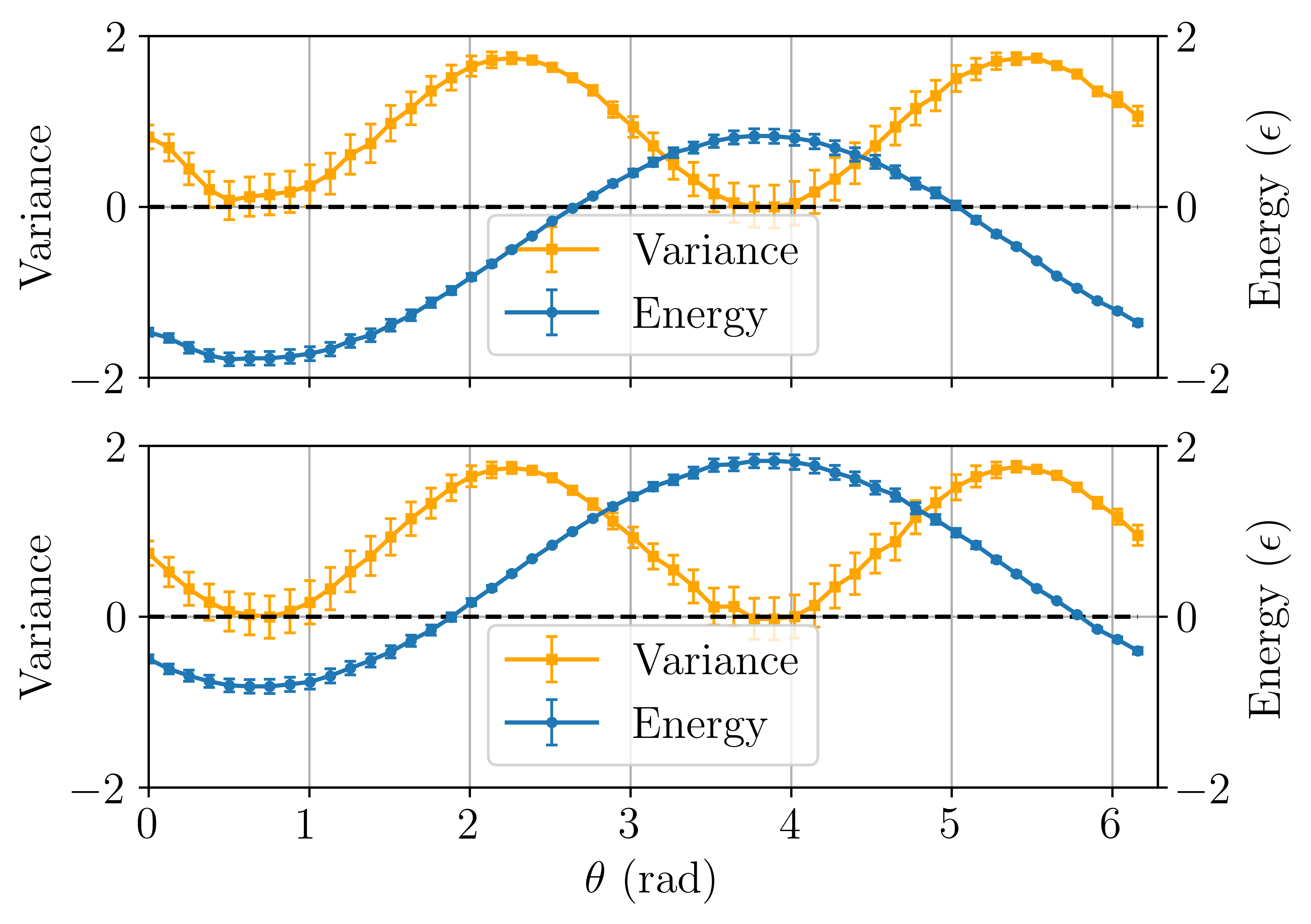}
    \caption{Parameter sweep of the block-diagonal submatrices, $A$ and $B$, for the $N=3$ LMG model, using 50 iteration steps, using quantum hardware.}
    \label{fig:sweep}
\end{figure}

\begin{figure}[tbh]
    \centering
    \includegraphics[width=8.6cm,clip]{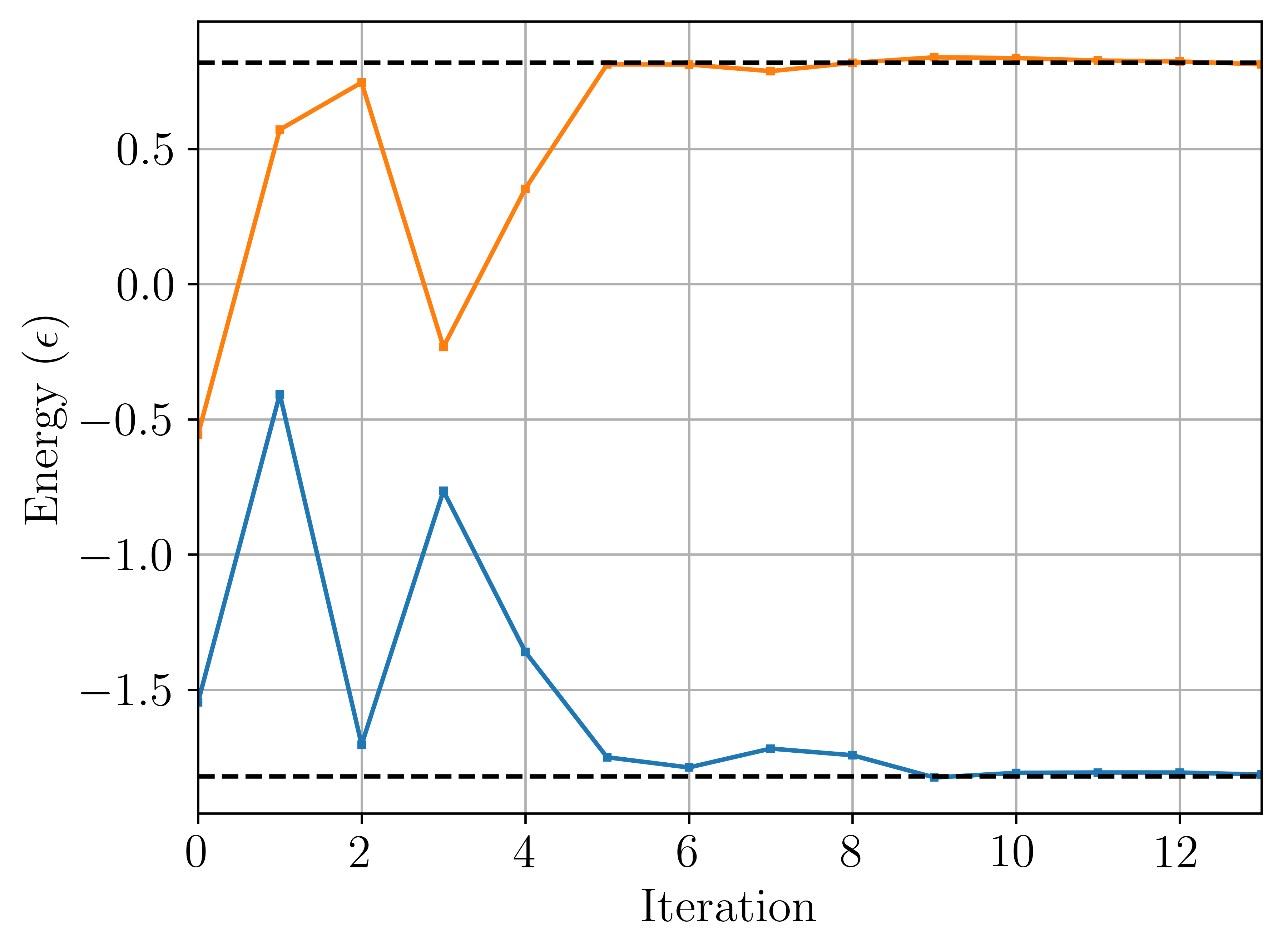}
    \caption{Iterative variance minimisation to converge on energy eigenstates (dashed lines) of the $H_{N=3A}$ sector of the $N=3$ LMG model with parameters $\frac{V}{\epsilon}=0.5$, $\frac{W}{\epsilon}=0$, using quantum hardware.  Random starting parameters are used, and two illustrative results are shown, resulting in the full spectrum.}
    \label{fig:LMG3_qc}
\end{figure}

We present the results of the calculations in Fig. \ref{fig:sweep}, which shows the parameter sweep, in Fig. \ref{fig:LMG3_qc} which shows the variance minimisations, and in Table \ref{tab:N=3_BD_results}, which shows the resulting states for which the Hamiltonian variance is zero. These results obtained on a quantum computer are close to the exact values for the respective eigenstates, within error.

\begin{table}[tbh]
\caption{\label{tab:N=3_BD_results}%
$N=3$ LMG model block-diagonal submatrix results calculated using the IBMQ\_manila quantum computer using the circuit in Fig. \ref{fig:1q_circuit_ansatz} using 20000 shots. The Ground and second excited state results are achieved using submatrix A, with submatrix B used for the first and third excited states.
}
\begin{ruledtabular}
\begin{tabular}{cdcc}
Eigenstate&
\multicolumn{1}{c}{\textrm{Exact Value ($\epsilon$)}}&
Variance&
\multicolumn{1}{c}{\textrm{QC Result ($\epsilon$)}}\\
\hline
Ground & -1.823 & 0.073 &-1.788 $\pm$ 0.062  \\
2nd & 0.823 & 0.001 &0.826 $\pm$ 0.064\\
\hline
1st & -0.823 & -0.004 & -0.816 $\pm$ 0.063\\
3rd & 1.823 & 0.001 & 1.810 $\pm$ 0.063\\
\end{tabular}
\end{ruledtabular}
\end{table}


\section{\label{N=7} \texorpdfstring{$N=7$}{N=7} Simulation and Measurement}

\subsection{Ansatz}
The $N=7$ submatrix featuring the lowest unperturbed eigenvalue in the diagonal for our choice of $V/\epsilon=0.5$, $W=0$ is 

 \begin{equation}
     H_{N=7 \text{ Submatrix}} = \begin{bmatrix}
     -3.5 & -2.291 & 0 & 0\\
     -2.291 & -1.5 & -3.873 & 0 \\
     0 & -3.873 & 0.5 & -3.354 \\
     0 & 0 & -3.354 & 2.5 \\
     \end{bmatrix}. \label{LMG7_matrix}
\end{equation}

 When encoded into the form of Pauli spin matrices, this becomes 

 \begin{align*}\label{eq:H7}
     H_{N=7} = &-0.5 - 2.8225X_1 -1.0Z_1 -1.9365X_0X_1\\
     &-1.9365Y_0Y_1 -2.0Z_0 + 0.5315Z_0X_1. \numberthis
 \end{align*}

As with the $N=3$ case the eigenvalues of the opposite parity submatrix can be obtained from those of the above matrix through multiplication by $-1$.

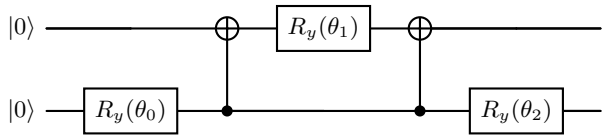
\begin{figure}[tbh]
			       \centering
			       \begin{quantikz}
				\lstick{\ket{0}} & \qw 
                      & \targ{} & \gate{R_y(\theta_1)} & \targ{}&\qw &\qw\\
                \lstick{\ket{0}} & \gate{R_y(\theta_0)} & \ctrl{-1}&\qw & \ctrl{-1} & \gate{R_y(\theta_2)} &\qw\\
				\end{quantikz}
			       \caption{Quantum circuit to calculate the eigenvalues of LMG model $N=7$ Hamiltonian.}
			       \label{fig:Lip_N=7}
			   \end{figure}

From Eq. (\ref{eq:H7}), a two-qubit circuit is required to represent the wave function of the system.  A sufficiently general circuit for this case, with three parameters, is shown in Fig. \ref{fig:Lip_N=7}.

For the calculation of the variance, the square of $H_{N=7}$ is needed.  Matrix multiplication gives

 \begin{equation}
     H^2_{N=7} = \begin{bmatrix}
 17.498681 & 11.455    &  8.873043&   0.      \\
 11.455    & 22.49881  &  3.873   &  12.990042\\
  8.873043 &  3.873    & 26.499445& -10.062   \\
  0.       & 12.990042 &-10.062   &  17.499316\\
     \end{bmatrix}, \label{LMG7_matrix_sq}
\end{equation}

\noindent from which Pauli decomposition produces the form needed for quantum computation as
\begin{align*} \label{eq:H7sq}
H^2_{N=7} =& 20.999063 + 0.6965 X_1 + 1.0 Z_1 + 10.9315425 X_0 
\\&+ 1.9365X_0X_1 -2.0584995X_0Z_1 + 1.93654 Y_0Y_1 \\&-1.0003175 Z_0 + 10.7585 Z_0X_1 -3.5 Z_0Z_1. \numberthis
\end{align*}

\subsection{Simulation}

\begin{figure}[tbh]
    \centering
    \includegraphics[width=8.6cm,clip]{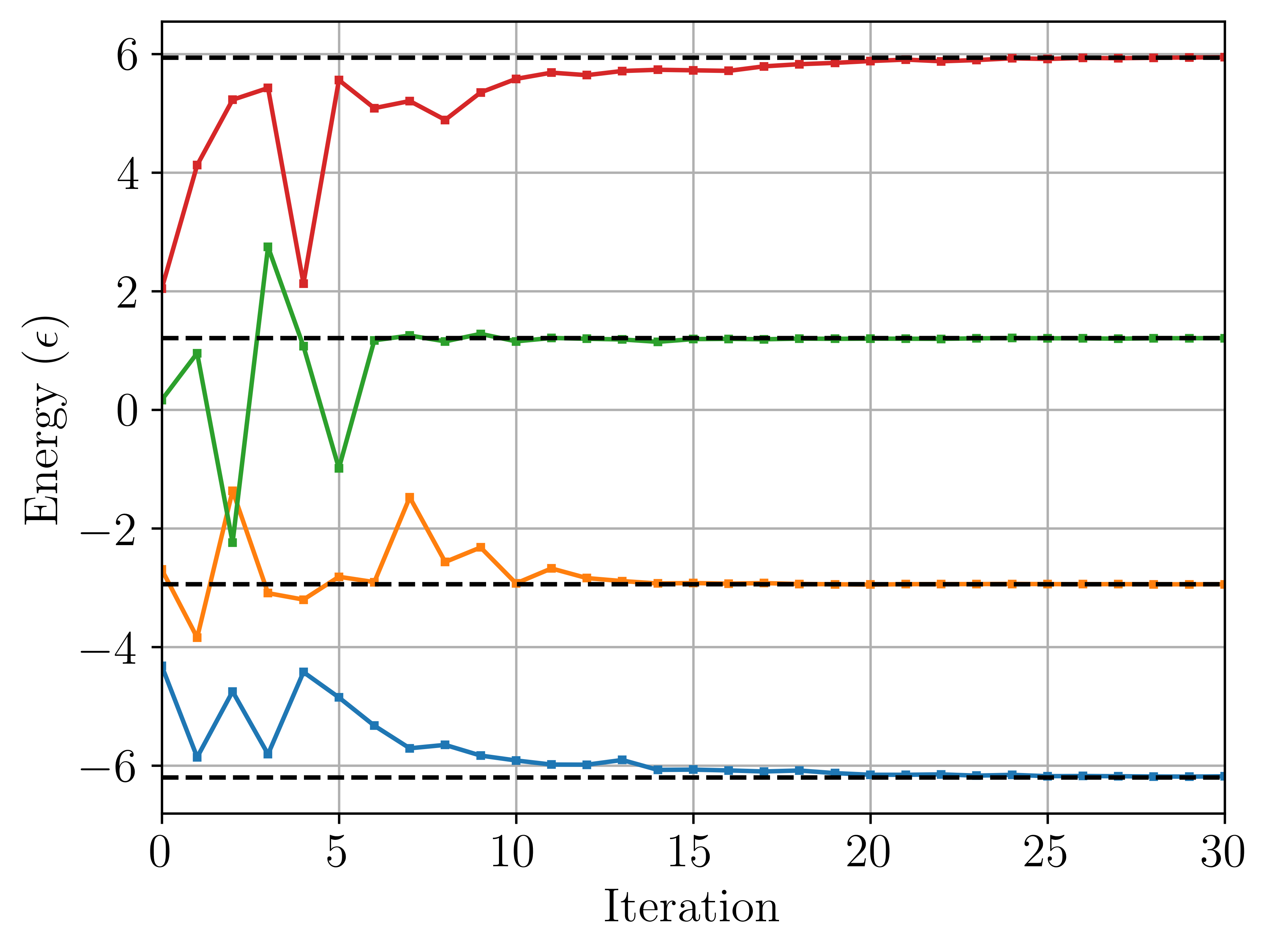}
    \caption{Energy values during iteration of the simulated variance minimisation method for the $N=7$ LMG model with parameters $\frac{V}{\epsilon}=0.5$, $\frac{W}{\epsilon}=0$.  Known exact eigenvalues are shown with dashed lines.  Four random starting points are shown corresponding to runs which find the four eigenstates.}
    \label{fig:LMG7_sim}
\end{figure}

Similarly to the $N=3$ simulation in Section \ref{sec:LMG3_simulation}, we perform noiseless simulations of the variance minimisation to find each eigenvalue. The results of these simulations for the $N=7$ LMG model are plotted in Fig. \ref{fig:LMG7_sim}, which shows convergence across each of the four simulations to their respective exact eigenstate. These simulations are also started from randomised initial parameters.

\subsection{Quantum Hardware Results}

\begin{figure}[tbh]
    \centering
    \includegraphics[width=8.6cm,clip]{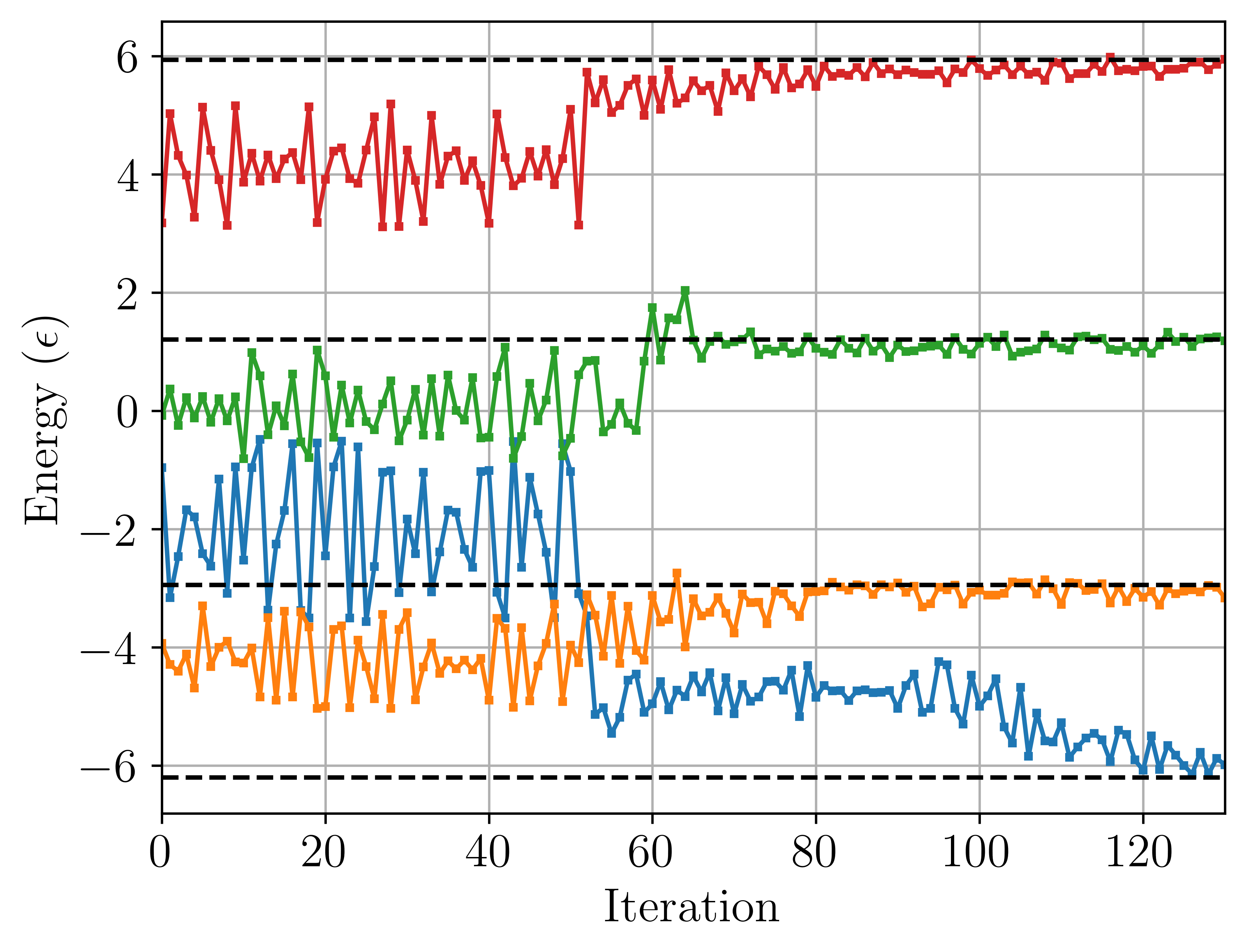}
    \caption{Energy values during iteration of the variance minimisation method on quantum hardware for the $N=7$ LMG model with parameters $\frac{V}{\epsilon}=0.5$, $\frac{W}{\epsilon}=0$. Known exact eigenvalues are shown with dashed lines.  Four random starting points are shown corresponding to runs which find the four eigenstates.}
    \label{fig:LMG7_qc}
\end{figure}

Measurements for the $N=7$ LMG excitation spectra were performed using the IBM\_nairobi quantum computer, again using 20,000 shots per circuit. We apply the same method of readout error mitigation as with the $N=3$ case, described in Section \ref{sec:Quantum Hardware Results}, as well as applying a method of CNOT mitigation. This mitigation technique adds pairs of CNOT gates to the circuit where there is a CNOT present, allowing for a linear extrapolation to the theoretical case where there are no CNOT gates in the circuit. Adding additional pairs of CNOT gates does not alter the value of the measurement, other than introducing additional error from the added CNOT pairs \cite{PhysRevX.7.021050}. The results of the iterative variance minimizations on IBM\_nairobi are presented in Fig. \ref{fig:LMG7_qc} and in Table \ref{tab:N=7_results}. 
The results are close to the known, exact solutions. 
However, these results are further from their respective exact solutions than the $N=3$ results. This is likely due to the increase in circuit depth required to perform the calculations for the $N=7$ case, as well as the additional qubit, resulting in greater gate error and noise within the circuit.  
 

\begin{table}[tbh]
\caption{\label{tab:N=7_results}%
$N=7$ LMG model results calculated using the IBMQ\_manila quantum computer using 20000 shots.
}
\begin{ruledtabular}
\begin{tabular}{cdcc}
Eigenstate&
\multicolumn{1}{c}{\textrm{Exact Value ($\epsilon$)}}&
Variance&
\multicolumn{1}{c}{\textrm{QC Result ($\epsilon$)}}\\
\hline
Ground & -6.208 & 0.139 &-6.067 $\pm$ 0.901\\
1st & -2.944 & 0.016 &-3.151 $\pm$ 0.503  \\
2nd & 1.208 & 0.010 &1.184 $\pm$ 0.484\\
3rd & 5.944 & 0.114 &5.902 $\pm$ 0.660\\
\end{tabular}
\end{ruledtabular}
\end{table}

In order to explore further the quality of the solutions obtained we calculate the fidelity, or overlap of the wavefunctions obtained by the quantum computer with the known analytic wavefunctions for each eigenvalue of the $N=7$ LMG model. These overlaps are given by 

\begin{equation}
    \langle \psi_{A} | \psi_{QC} \rangle ^2
\end{equation}

\noindent where $\psi_A$ is the wavefunction of the analytic solution, and $\psi_{QC}$ is the wavefunction determined by the quantum computer. The results of the overlap  calculations are shown in Table \ref{tab:N=7_Projections}. 

\begin{table}[tbh]
\caption{\label{tab:N=7_Projections}%
Overlaps of the $N=7$ LMG model result wavefunctions with analytic solution wavefunctions.}
\begin{ruledtabular}
\begin{tabular}{cdddd}
Measured Eigenstate &-6.2 & -2.94 & 1.2 & 5.94\\
\hline
Ground & 0.924 & 0.073 & 0.002 & 0.001\\
1st & 0.020 & 0.947 & 0.033 & <0.001 \\
2nd & 0.001 & 0.009 & 0.991 & <0.001\\
3rd & 0.001 & <0.001 & 0.026 & 0.972\\
\end{tabular}
\end{ruledtabular}
\end{table}

 Here we see that all results found using variance minimization on quantum hardware find states that closely overlap with their respective known analytic eigenvector. The ground state introduces slightly more overlap with the first excited state, which could be due to only achieving a partial convergence in the time available for the run on real quantum hardware, as can be seen in Fig. \ref{fig:LMG7_qc}.

\section{Conclusion}
We adapted the Variational Quantum Eigensolver (VQE) to target any eigenstate of a Hamiltonian through minimization of the variance of the Hamiltonian.  The cost is in the form of increased circuit evaluations compared with a ground state VQE, but not circuit depth or number of qubits.  States close to zero variance could be found on presently-available quantum computing hardware, corresponding to the known eigenstates of a model Hamiltonian.  Further exploration of larger problems would be needed to check how reliably one could extract a full spectrum with the necessarily increasingly complex circuits needed to parameterize larger problems. 
\begin{acknowledgments}
This work was funded by AWE. We acknowledge the use of IBM Quantum services for this work. The views expressed are those of the authors, and do not reflect the official policy or position of IBM or the IBM Quantum team. In this paper we used IBMQ\_manilla and IBMQ\_nairobi, which are IBM Quantum Processors.  PDS acknowledges support from the UK STFC under grant numbers ST/V001108/1 and ST/W006472/1. UK Ministry of Defence \copyright  Crown owned copyright 2024/AWE.
\end{acknowledgments}


\bibliography{biblio.bib}

\begin{thebibliography}{41}%
\makeatletter
\providecommand \@ifxundefined [1]{%
 \@ifx{#1\undefined}
}%
\providecommand \@ifnum [1]{%
 \ifnum #1\expandafter \@firstoftwo
 \else \expandafter \@secondoftwo
 \fi
}%
\providecommand \@ifx [1]{%
 \ifx #1\expandafter \@firstoftwo
 \else \expandafter \@secondoftwo
 \fi
}%
\providecommand \natexlab [1]{#1}%
\providecommand \enquote  [1]{``#1''}%
\providecommand \bibnamefont  [1]{#1}%
\providecommand \bibfnamefont [1]{#1}%
\providecommand \citenamefont [1]{#1}%
\providecommand \href@noop [0]{\@secondoftwo}%
\providecommand \href [0]{\begingroup \@sanitize@url \@href}%
\providecommand \@href[1]{\@@startlink{#1}\@@href}%
\providecommand \@@href[1]{\endgroup#1\@@endlink}%
\providecommand \@sanitize@url [0]{\catcode `\\12\catcode `\$12\catcode `\&12\catcode `\#12\catcode `\^12\catcode `\_12\catcode `\%12\relax}%
\providecommand \@@startlink[1]{}%
\providecommand \@@endlink[0]{}%
\providecommand \url  [0]{\begingroup\@sanitize@url \@url }%
\providecommand \@url [1]{\endgroup\@href {#1}{\urlprefix }}%
\providecommand \urlprefix  [0]{URL }%
\providecommand \Eprint [0]{\href }%
\providecommand \doibase [0]{https://doi.org/}%
\providecommand \selectlanguage [0]{\@gobble}%
\providecommand \bibinfo  [0]{\@secondoftwo}%
\providecommand \bibfield  [0]{\@secondoftwo}%
\providecommand \translation [1]{[#1]}%
\providecommand \BibitemOpen [0]{}%
\providecommand \bibitemStop [0]{}%
\providecommand \bibitemNoStop [0]{.\EOS\space}%
\providecommand \EOS [0]{\spacefactor3000\relax}%
\providecommand \BibitemShut  [1]{\csname bibitem#1\endcsname}%
\let\auto@bib@innerbib\@empty
\bibitem [{\citenamefont {Vi{\v s}{\v n}{\'a}k}(2015)}]{visnak_quantum_2015}%
  \BibitemOpen
  \bibfield  {author} {\bibinfo {author} {\bibfnamefont {J.}~\bibnamefont {Vi{\v s}{\v n}{\'a}k}},\ }\bibfield  {title} {\bibinfo {title} {Quantum algorithms for computational nuclear physics},\ }\href {https://doi.org/10.1051/epjconf/201510001008} {\bibfield  {journal} {\bibinfo  {journal} {EPJ Web of Conferences}\ }\textbf {\bibinfo {volume} {100}},\ \bibinfo {pages} {01008} (\bibinfo {year} {2015})}\BibitemShut {NoStop}%
\bibitem [{\citenamefont {Vi{\v s}{\v n}{\'a}k}\ and\ \citenamefont {Vesel{\'y}}(2017)}]{visnak_quantum_2017}%
  \BibitemOpen
  \bibfield  {author} {\bibinfo {author} {\bibfnamefont {J.}~\bibnamefont {Vi{\v s}{\v n}{\'a}k}}\ and\ \bibinfo {author} {\bibfnamefont {P.}~\bibnamefont {Vesel{\'y}}},\ }\bibfield  {title} {\bibinfo {title} {Quantum algorithms for computational nuclear physics revisited, particular case of second quantized formulation},\ }\href {https://doi.org/10.1051/epjconf/201715401030} {\bibfield  {journal} {\bibinfo  {journal} {EPJ Web of Conferences}\ }\textbf {\bibinfo {volume} {154}},\ \bibinfo {pages} {01030} (\bibinfo {year} {2017})}\BibitemShut {NoStop}%
\bibitem [{\citenamefont {Dumitrescu}\ \emph {et~al.}(2018)\citenamefont {Dumitrescu}, \citenamefont {McCaskey}, \citenamefont {Hagen}, \citenamefont {Jansen}, \citenamefont {Morris}, \citenamefont {Papenbrock}, \citenamefont {Pooser}, \citenamefont {Dean},\ and\ \citenamefont {Lougovski}}]{dumitrescu_cloud_2018}%
  \BibitemOpen
  \bibfield  {author} {\bibinfo {author} {\bibfnamefont {E.~F.}\ \bibnamefont {Dumitrescu}}, \bibinfo {author} {\bibfnamefont {A.~J.}\ \bibnamefont {McCaskey}}, \bibinfo {author} {\bibfnamefont {G.}~\bibnamefont {Hagen}}, \bibinfo {author} {\bibfnamefont {G.~R.}\ \bibnamefont {Jansen}}, \bibinfo {author} {\bibfnamefont {T.~D.}\ \bibnamefont {Morris}}, \bibinfo {author} {\bibfnamefont {T.}~\bibnamefont {Papenbrock}}, \bibinfo {author} {\bibfnamefont {R.~C.}\ \bibnamefont {Pooser}}, \bibinfo {author} {\bibfnamefont {D.~J.}\ \bibnamefont {Dean}},\ and\ \bibinfo {author} {\bibfnamefont {P.}~\bibnamefont {Lougovski}},\ }\bibfield  {title} {\bibinfo {title} {Cloud {Quantum} {Computing} of an {Atomic} {Nucleus}},\ }\href {https://doi.org/10.1103/PhysRevLett.120.210501} {\bibfield  {journal} {\bibinfo  {journal} {Physical Review Letters}\ }\textbf {\bibinfo {volume} {120}},\ \bibinfo {pages} {210501} (\bibinfo {year} {2018})}\BibitemShut {NoStop}%
\bibitem [{\citenamefont {Roggero}\ \emph {et~al.}(2020{\natexlab{a}})\citenamefont {Roggero}, \citenamefont {Gu}, \citenamefont {Baroni},\ and\ \citenamefont {Papenbrock}}]{roggero_preparation_2020}%
  \BibitemOpen
  \bibfield  {author} {\bibinfo {author} {\bibfnamefont {A.}~\bibnamefont {Roggero}}, \bibinfo {author} {\bibfnamefont {C.}~\bibnamefont {Gu}}, \bibinfo {author} {\bibfnamefont {A.}~\bibnamefont {Baroni}},\ and\ \bibinfo {author} {\bibfnamefont {T.}~\bibnamefont {Papenbrock}},\ }\bibfield  {title} {\bibinfo {title} {Preparation of excited states for nuclear dynamics on a quantum computer},\ }\href {https://doi.org/10.1103/PhysRevC.102.064624} {\bibfield  {journal} {\bibinfo  {journal} {Physical Review C}\ }\textbf {\bibinfo {volume} {102}},\ \bibinfo {pages} {1} (\bibinfo {year} {2020}{\natexlab{a}})}\BibitemShut {NoStop}%
\bibitem [{\citenamefont {Roggero}\ \emph {et~al.}(2020{\natexlab{b}})\citenamefont {Roggero}, \citenamefont {Li}, \citenamefont {Carlson}, \citenamefont {Gupta},\ and\ \citenamefont {Perdue}}]{roggero_quantum_2020}%
  \BibitemOpen
  \bibfield  {author} {\bibinfo {author} {\bibfnamefont {A.}~\bibnamefont {Roggero}}, \bibinfo {author} {\bibfnamefont {A.~C.}\ \bibnamefont {Li}}, \bibinfo {author} {\bibfnamefont {J.}~\bibnamefont {Carlson}}, \bibinfo {author} {\bibfnamefont {R.}~\bibnamefont {Gupta}},\ and\ \bibinfo {author} {\bibfnamefont {G.~N.}\ \bibnamefont {Perdue}},\ }\bibfield  {title} {\bibinfo {title} {Quantum computing for neutrino-nucleus scattering},\ }\href {https://doi.org/10.1103/PhysRevD.101.074038} {\bibfield  {journal} {\bibinfo  {journal} {Physical Review D}\ }\textbf {\bibinfo {volume} {101}},\ \bibinfo {pages} {74038} (\bibinfo {year} {2020}{\natexlab{b}})}\BibitemShut {NoStop}%
\bibitem [{\citenamefont {Lacroix}(2020)}]{lacroix_symmetry-assisted_2020}%
  \BibitemOpen
  \bibfield  {author} {\bibinfo {author} {\bibfnamefont {D.}~\bibnamefont {Lacroix}},\ }\bibfield  {title} {\bibinfo {title} {Symmetry-{Assisted} {Preparation} of {Entangled} {Many}-{Body} {States} on a {Quantum} {Computer}},\ }\href {https://doi.org/10.1103/PhysRevLett.125.230502} {\bibfield  {journal} {\bibinfo  {journal} {Physical Review Letters}\ }\textbf {\bibinfo {volume} {125}},\ \bibinfo {pages} {230502} (\bibinfo {year} {2020})}\BibitemShut {NoStop}%
\bibitem [{\citenamefont {Siwach}\ and\ \citenamefont {Lacroix}(2021)}]{siwach_filtering_2021}%
  \BibitemOpen
  \bibfield  {author} {\bibinfo {author} {\bibfnamefont {P.}~\bibnamefont {Siwach}}\ and\ \bibinfo {author} {\bibfnamefont {D.}~\bibnamefont {Lacroix}},\ }\bibfield  {title} {\bibinfo {title} {Filtering states with total spin on a quantum computer},\ }\href {https://doi.org/10.1103/PhysRevA.104.062435} {\bibfield  {journal} {\bibinfo  {journal} {Physical Review A}\ }\textbf {\bibinfo {volume} {104}},\ \bibinfo {pages} {062435} (\bibinfo {year} {2021})}\BibitemShut {NoStop}%
\bibitem [{\citenamefont {Baroni}\ \emph {et~al.}(2022)\citenamefont {Baroni}, \citenamefont {Carlson}, \citenamefont {Gupta}, \citenamefont {Li}, \citenamefont {Perdue},\ and\ \citenamefont {Roggero}}]{baroni_nuclear_2021}%
  \BibitemOpen
  \bibfield  {author} {\bibinfo {author} {\bibfnamefont {A.}~\bibnamefont {Baroni}}, \bibinfo {author} {\bibfnamefont {J.}~\bibnamefont {Carlson}}, \bibinfo {author} {\bibfnamefont {R.}~\bibnamefont {Gupta}}, \bibinfo {author} {\bibfnamefont {A.~C.~Y.}\ \bibnamefont {Li}}, \bibinfo {author} {\bibfnamefont {G.~N.}\ \bibnamefont {Perdue}},\ and\ \bibinfo {author} {\bibfnamefont {A.}~\bibnamefont {Roggero}},\ }\bibfield  {title} {\bibinfo {title} {Nuclear two point correlation functions on a quantum computer},\ }\href {https://doi.org/10.1103/PhysRevD.105.074503} {\bibfield  {journal} {\bibinfo  {journal} {Phys. Rev. D}\ }\textbf {\bibinfo {volume} {105}},\ \bibinfo {pages} {074503} (\bibinfo {year} {2022})}\BibitemShut {NoStop}%
\bibitem [{\citenamefont {Robin}\ \emph {et~al.}(2021)\citenamefont {Robin}, \citenamefont {Savage},\ and\ \citenamefont {Pillet}}]{robin_entanglement_2021}%
  \BibitemOpen
  \bibfield  {author} {\bibinfo {author} {\bibfnamefont {C.}~\bibnamefont {Robin}}, \bibinfo {author} {\bibfnamefont {M.~J.}\ \bibnamefont {Savage}},\ and\ \bibinfo {author} {\bibfnamefont {N.}~\bibnamefont {Pillet}},\ }\bibfield  {title} {\bibinfo {title} {Entanglement rearrangement in self-consistent nuclear structure calculations},\ }\href {https://doi.org/10.1103/PhysRevC.103.034325} {\bibfield  {journal} {\bibinfo  {journal} {Physical Review C}\ }\textbf {\bibinfo {volume} {103}},\ \bibinfo {pages} {034325} (\bibinfo {year} {2021})}\BibitemShut {NoStop}%
\bibitem [{\citenamefont {Zhang}\ \emph {et~al.}(2021)\citenamefont {Zhang}, \citenamefont {Xing}, \citenamefont {Yan}, \citenamefont {Wang},\ and\ \citenamefont {Zhu}}]{zhang_selected_2021}%
  \BibitemOpen
  \bibfield  {author} {\bibinfo {author} {\bibfnamefont {D.-B.}\ \bibnamefont {Zhang}}, \bibinfo {author} {\bibfnamefont {H.}~\bibnamefont {Xing}}, \bibinfo {author} {\bibfnamefont {H.}~\bibnamefont {Yan}}, \bibinfo {author} {\bibfnamefont {E.}~\bibnamefont {Wang}},\ and\ \bibinfo {author} {\bibfnamefont {S.-L.}\ \bibnamefont {Zhu}},\ }\bibfield  {title} {\bibinfo {title} {Selected topics of quantum computing for nuclear physics},\ }\href {https://doi.org/10.1088/1674-1056/abd761} {\bibfield  {journal} {\bibinfo  {journal} {Chinese Physics B}\ }\textbf {\bibinfo {volume} {30}},\ \bibinfo {pages} {020306} (\bibinfo {year} {2021})}\BibitemShut {NoStop}%
\bibitem [{\citenamefont {Robbins}\ and\ \citenamefont {Love}(2021)}]{robbins_benchmarking_2021}%
  \BibitemOpen
  \bibfield  {author} {\bibinfo {author} {\bibfnamefont {K.}~\bibnamefont {Robbins}}\ and\ \bibinfo {author} {\bibfnamefont {P.~J.}\ \bibnamefont {Love}},\ }\bibfield  {title} {\bibinfo {title} {Benchmarking near-term quantum devices with the variational quantum eigensolver and the lipkin-meshkov-glick model},\ }\href {https://doi.org/10.1103/PhysRevA.104.022412} {\bibfield  {journal} {\bibinfo  {journal} {Phys. Rev. A}\ }\textbf {\bibinfo {volume} {104}},\ \bibinfo {pages} {022412} (\bibinfo {year} {2021})}\BibitemShut {NoStop}%
\bibitem [{\citenamefont {P{\'e}rez-Fern{\'a}ndez}\ \emph {et~al.}(2022)\citenamefont {P{\'e}rez-Fern{\'a}ndez}, \citenamefont {Arias}, \citenamefont {Garc{\'i}a-Ramos},\ and\ \citenamefont {Lamata}}]{perez-fernandez_digital_2022}%
  \BibitemOpen
  \bibfield  {author} {\bibinfo {author} {\bibfnamefont {P.}~\bibnamefont {P{\'e}rez-Fern{\'a}ndez}}, \bibinfo {author} {\bibfnamefont {J.-M.}\ \bibnamefont {Arias}}, \bibinfo {author} {\bibfnamefont {J.-E.}\ \bibnamefont {Garc{\'i}a-Ramos}},\ and\ \bibinfo {author} {\bibfnamefont {L.}~\bibnamefont {Lamata}},\ }\bibfield  {title} {\bibinfo {title} {A digital quantum simulation of the {Agassi} model},\ }\href {https://doi.org/10.1016/j.physletb.2022.137133} {\bibfield  {journal} {\bibinfo  {journal} {Physics Letters B}\ ,\ \bibinfo {pages} {137133}} (\bibinfo {year} {2022})}\BibitemShut {NoStop}%
\bibitem [{\citenamefont {Cervia}\ \emph {et~al.}(2021)\citenamefont {Cervia}, \citenamefont {Balantekin}, \citenamefont {Coppersmith}, \citenamefont {Johnson}, \citenamefont {Love}, \citenamefont {Poole}, \citenamefont {Robbins},\ and\ \citenamefont {Saffman}}]{cervia_lipkin_2021}%
  \BibitemOpen
  \bibfield  {author} {\bibinfo {author} {\bibfnamefont {M.~J.}\ \bibnamefont {Cervia}}, \bibinfo {author} {\bibfnamefont {A.~B.}\ \bibnamefont {Balantekin}}, \bibinfo {author} {\bibfnamefont {S.~N.}\ \bibnamefont {Coppersmith}}, \bibinfo {author} {\bibfnamefont {C.~W.}\ \bibnamefont {Johnson}}, \bibinfo {author} {\bibfnamefont {P.~J.}\ \bibnamefont {Love}}, \bibinfo {author} {\bibfnamefont {C.}~\bibnamefont {Poole}}, \bibinfo {author} {\bibfnamefont {K.}~\bibnamefont {Robbins}},\ and\ \bibinfo {author} {\bibfnamefont {M.}~\bibnamefont {Saffman}},\ }\bibfield  {title} {\bibinfo {title} {Lipkin model on a quantum computer},\ }\href {https://doi.org/10.1103/PhysRevC.104.024305} {\bibfield  {journal} {\bibinfo  {journal} {Physical Review C}\ }\textbf {\bibinfo {volume} {104}},\ \bibinfo {pages} {024305} (\bibinfo {year} {2021})}\BibitemShut {NoStop}%
\bibitem [{\citenamefont {Kiss}\ \emph {et~al.}(2022)\citenamefont {Kiss}, \citenamefont {Grossi}, \citenamefont {Lougovski}, \citenamefont {Sanchez}, \citenamefont {Vallecorsa},\ and\ \citenamefont {Papenbrock}}]{kiss_quantum_2022}%
  \BibitemOpen
  \bibfield  {author} {\bibinfo {author} {\bibfnamefont {O.}~\bibnamefont {Kiss}}, \bibinfo {author} {\bibfnamefont {M.}~\bibnamefont {Grossi}}, \bibinfo {author} {\bibfnamefont {P.}~\bibnamefont {Lougovski}}, \bibinfo {author} {\bibfnamefont {F.}~\bibnamefont {Sanchez}}, \bibinfo {author} {\bibfnamefont {S.}~\bibnamefont {Vallecorsa}},\ and\ \bibinfo {author} {\bibfnamefont {T.}~\bibnamefont {Papenbrock}},\ }\bibfield  {title} {\bibinfo {title} {Quantum computing of the $^6${L}i nucleus via ordered unitary coupled clusters},\ }\href {https://doi.org/10.1103/PhysRevC.106.034325} {\bibfield  {journal} {\bibinfo  {journal} {Physical Review C}\ }\textbf {\bibinfo {volume} {106}},\ \bibinfo {pages} {034325} (\bibinfo {year} {2022})}\BibitemShut {NoStop}%
\bibitem [{\citenamefont {Romero}\ \emph {et~al.}(2022)\citenamefont {Romero}, \citenamefont {Engel}, \citenamefont {Tang},\ and\ \citenamefont {Economou}}]{romero_solving_2022}%
  \BibitemOpen
  \bibfield  {author} {\bibinfo {author} {\bibfnamefont {A.~M.}\ \bibnamefont {Romero}}, \bibinfo {author} {\bibfnamefont {J.}~\bibnamefont {Engel}}, \bibinfo {author} {\bibfnamefont {H.~L.}\ \bibnamefont {Tang}},\ and\ \bibinfo {author} {\bibfnamefont {S.~E.}\ \bibnamefont {Economou}},\ }\bibfield  {title} {\bibinfo {title} {Solving nuclear structure problems with the adaptive variational quantum algorithm},\ }\href {https://doi.org/10.1103/PhysRevC.105.064317} {\bibfield  {journal} {\bibinfo  {journal} {Physical Review C}\ }\textbf {\bibinfo {volume} {105}},\ \bibinfo {pages} {064317} (\bibinfo {year} {2022})}\BibitemShut {NoStop}%
\bibitem [{\citenamefont {Chikaoka}\ and\ \citenamefont {Liang}(2022)}]{chikaoka_quantum_2022}%
  \BibitemOpen
  \bibfield  {author} {\bibinfo {author} {\bibfnamefont {A.}~\bibnamefont {Chikaoka}}\ and\ \bibinfo {author} {\bibfnamefont {H.}~\bibnamefont {Liang}},\ }\bibfield  {title} {\bibinfo {title} {Quantum computing for the {Lipkin} model with unitary coupled cluster and structure learning ansatz},\ }\href {https://doi.org/10.1088/1674-1137/ac380a} {\bibfield  {journal} {\bibinfo  {journal} {Chinese Physics C}\ }\textbf {\bibinfo {volume} {46}},\ \bibinfo {pages} {024106} (\bibinfo {year} {2022})}\BibitemShut {NoStop}%
\bibitem [{\citenamefont {Hlatshwayo}\ \emph {et~al.}(2022)\citenamefont {Hlatshwayo}, \citenamefont {Zhang}, \citenamefont {Wibowo}, \citenamefont {LaRose}, \citenamefont {Lacroix},\ and\ \citenamefont {Litvinova}}]{hlatshwayo_simulating_2022}%
  \BibitemOpen
  \bibfield  {author} {\bibinfo {author} {\bibfnamefont {M.~Q.}\ \bibnamefont {Hlatshwayo}}, \bibinfo {author} {\bibfnamefont {Y.}~\bibnamefont {Zhang}}, \bibinfo {author} {\bibfnamefont {H.}~\bibnamefont {Wibowo}}, \bibinfo {author} {\bibfnamefont {R.}~\bibnamefont {LaRose}}, \bibinfo {author} {\bibfnamefont {D.}~\bibnamefont {Lacroix}},\ and\ \bibinfo {author} {\bibfnamefont {E.}~\bibnamefont {Litvinova}},\ }\bibfield  {title} {\bibinfo {title} {Simulating excited states of the {Lipkin} model on a quantum computer},\ }\href {https://doi.org/10.1103/PhysRevC.106.024319} {\bibfield  {journal} {\bibinfo  {journal} {Physical Review C}\ }\textbf {\bibinfo {volume} {106}},\ \bibinfo {pages} {024319} (\bibinfo {year} {2022})}\BibitemShut {NoStop}%
\bibitem [{\citenamefont {Ruiz~Guzman}\ and\ \citenamefont {Lacroix}(2022)}]{ruiz_guzman_accessing_2022}%
  \BibitemOpen
  \bibfield  {author} {\bibinfo {author} {\bibfnamefont {E.~A.}\ \bibnamefont {Ruiz~Guzman}}\ and\ \bibinfo {author} {\bibfnamefont {D.}~\bibnamefont {Lacroix}},\ }\bibfield  {title} {\bibinfo {title} {Accessing ground-state and excited-state energies in a many-body system after symmetry restoration using quantum computers},\ }\href {https://doi.org/10.1103/PhysRevC.105.024324} {\bibfield  {journal} {\bibinfo  {journal} {Physical Review C}\ }\textbf {\bibinfo {volume} {105}},\ \bibinfo {pages} {024324} (\bibinfo {year} {2022})}\BibitemShut {NoStop}%
\bibitem [{\citenamefont {Hobday}\ \emph {et~al.}(2022)\citenamefont {Hobday}, \citenamefont {Stevenson},\ and\ \citenamefont {Benstead}}]{hobday}%
  \BibitemOpen
  \bibfield  {author} {\bibinfo {author} {\bibfnamefont {I.}~\bibnamefont {Hobday}}, \bibinfo {author} {\bibfnamefont {P.~D.}\ \bibnamefont {Stevenson}},\ and\ \bibinfo {author} {\bibfnamefont {J.}~\bibnamefont {Benstead}},\ }\bibfield  {title} {\bibinfo {title} {{Quantum computing calculations for nuclear structure and nuclear data}},\ }in\ \href {https://doi.org/10.1117/12.2632782} {\emph {\bibinfo {booktitle} {Quantum Technologies 2022}}},\ Vol.\ \bibinfo {volume} {12133},\ \bibinfo {organization} {International Society for Optics and Photonics}\ (\bibinfo  {publisher} {SPIE},\ \bibinfo {year} {2022})\ p.\ \bibinfo {pages} {109}\BibitemShut {NoStop}%
\bibitem [{\citenamefont {Stetcu}\ \emph {et~al.}(2022)\citenamefont {Stetcu}, \citenamefont {Baroni},\ and\ \citenamefont {Carlson}}]{PhysRevC.105.064308}%
  \BibitemOpen
  \bibfield  {author} {\bibinfo {author} {\bibfnamefont {I.}~\bibnamefont {Stetcu}}, \bibinfo {author} {\bibfnamefont {A.}~\bibnamefont {Baroni}},\ and\ \bibinfo {author} {\bibfnamefont {J.}~\bibnamefont {Carlson}},\ }\bibfield  {title} {\bibinfo {title} {Variational approaches to constructing the many-body nuclear ground state for quantum computing},\ }\href {https://doi.org/10.1103/PhysRevC.105.064308} {\bibfield  {journal} {\bibinfo  {journal} {Phys. Rev. C}\ }\textbf {\bibinfo {volume} {105}},\ \bibinfo {pages} {064308} (\bibinfo {year} {2022})}\BibitemShut {NoStop}%
\bibitem [{\citenamefont {Li}\ \emph {et~al.}(2023)\citenamefont {Li}, \citenamefont {Al-Khalili},\ and\ \citenamefont {Stevenson}}]{li2023quantum}%
  \BibitemOpen
  \bibfield  {author} {\bibinfo {author} {\bibfnamefont {Y.~H.}\ \bibnamefont {Li}}, \bibinfo {author} {\bibfnamefont {J.}~\bibnamefont {Al-Khalili}},\ and\ \bibinfo {author} {\bibfnamefont {P.}~\bibnamefont {Stevenson}},\ }\href@noop {} {\bibinfo {title} {A quantum simulation approach to implementing nuclear density functional theory via imaginary time evolution}} (\bibinfo {year} {2023}),\ \Eprint {https://arxiv.org/abs/2308.15425} {arXiv:2308.15425} \BibitemShut {NoStop}%
\bibitem [{\citenamefont {Feynman}(1982)}]{feynman_simulating_1982}%
  \BibitemOpen
  \bibfield  {author} {\bibinfo {author} {\bibfnamefont {R.~P.}\ \bibnamefont {Feynman}},\ }\bibfield  {title} {\bibinfo {title} {Simulating physics with computers},\ }\href@noop {} {\bibfield  {journal} {\bibinfo  {journal} {International Journal of Theoretical Physics}\ }\textbf {\bibinfo {volume} {21}},\ \bibinfo {pages} {467} (\bibinfo {year} {1982})}\BibitemShut {NoStop}%
\bibitem [{\citenamefont {O’Brien}\ \emph {et~al.}(2019)\citenamefont {O’Brien}, \citenamefont {Tarasinski},\ and\ \citenamefont {Terhal}}]{OBrien_2019}%
  \BibitemOpen
  \bibfield  {author} {\bibinfo {author} {\bibfnamefont {T.~E.}\ \bibnamefont {O’Brien}}, \bibinfo {author} {\bibfnamefont {B.}~\bibnamefont {Tarasinski}},\ and\ \bibinfo {author} {\bibfnamefont {B.~M.}\ \bibnamefont {Terhal}},\ }\bibfield  {title} {\bibinfo {title} {Quantum phase estimation of multiple eigenvalues for small-scale (noisy) experiments},\ }\href {https://doi.org/10.1088/1367-2630/aafb8e} {\bibfield  {journal} {\bibinfo  {journal} {New Journal of Physics}\ }\textbf {\bibinfo {volume} {21}},\ \bibinfo {pages} {023022} (\bibinfo {year} {2019})}\BibitemShut {NoStop}%
\bibitem [{\citenamefont {Zhou}\ \emph {et~al.}(2013)\citenamefont {Zhou}, \citenamefont {Kalasuwan}, \citenamefont {Ralph},\ and\ \citenamefont {O'brien}}]{zhou2013calculating}%
  \BibitemOpen
  \bibfield  {author} {\bibinfo {author} {\bibfnamefont {X.-Q.}\ \bibnamefont {Zhou}}, \bibinfo {author} {\bibfnamefont {P.}~\bibnamefont {Kalasuwan}}, \bibinfo {author} {\bibfnamefont {T.~C.}\ \bibnamefont {Ralph}},\ and\ \bibinfo {author} {\bibfnamefont {J.~L.}\ \bibnamefont {O'brien}},\ }\bibfield  {title} {\bibinfo {title} {Calculating unknown eigenvalues with a quantum algorithm},\ }\href@noop {} {\bibfield  {journal} {\bibinfo  {journal} {Nature photonics}\ }\textbf {\bibinfo {volume} {7}},\ \bibinfo {pages} {223} (\bibinfo {year} {2013})}\BibitemShut {NoStop}%
\bibitem [{\citenamefont {Du}\ \emph {et~al.}(2021)\citenamefont {Du}, \citenamefont {Vary}, \citenamefont {Zhao},\ and\ \citenamefont {Zuo}}]{du2021ab}%
  \BibitemOpen
  \bibfield  {author} {\bibinfo {author} {\bibfnamefont {W.}~\bibnamefont {Du}}, \bibinfo {author} {\bibfnamefont {J.~P.}\ \bibnamefont {Vary}}, \bibinfo {author} {\bibfnamefont {X.}~\bibnamefont {Zhao}},\ and\ \bibinfo {author} {\bibfnamefont {W.}~\bibnamefont {Zuo}},\ }\href@noop {} {\bibinfo {title} {Ab initio nuclear structure via quantum adiabatic algorithm}} (\bibinfo {year} {2021}),\ \Eprint {https://arxiv.org/abs/2105.08910} {arXiv:2105.08910} \BibitemShut {NoStop}%
\bibitem [{\citenamefont {Motta}\ \emph {et~al.}(2020)\citenamefont {Motta}, \citenamefont {Sun}, \citenamefont {Tan}, \citenamefont {O'Rourke}, \citenamefont {Ye}, \citenamefont {Minnich}, \citenamefont {Brand{\~a}o},\ and\ \citenamefont {Chan}}]{Motta2020}%
  \BibitemOpen
  \bibfield  {author} {\bibinfo {author} {\bibfnamefont {M.}~\bibnamefont {Motta}}, \bibinfo {author} {\bibfnamefont {C.}~\bibnamefont {Sun}}, \bibinfo {author} {\bibfnamefont {A.~T.~K.}\ \bibnamefont {Tan}}, \bibinfo {author} {\bibfnamefont {M.~J.}\ \bibnamefont {O'Rourke}}, \bibinfo {author} {\bibfnamefont {E.}~\bibnamefont {Ye}}, \bibinfo {author} {\bibfnamefont {A.~J.}\ \bibnamefont {Minnich}}, \bibinfo {author} {\bibfnamefont {F.~G. S.~L.}\ \bibnamefont {Brand{\~a}o}},\ and\ \bibinfo {author} {\bibfnamefont {G.~K.-L.}\ \bibnamefont {Chan}},\ }\bibfield  {title} {\bibinfo {title} {Determining eigenstates and thermal states on a quantum computer using quantum imaginary time evolution},\ }\href {https://doi.org/10.1038/s41567-019-0704-4} {\bibfield  {journal} {\bibinfo  {journal} {Nature Physics}\ }\textbf {\bibinfo {volume} {16}},\ \bibinfo {pages} {205} (\bibinfo {year} {2020})}\BibitemShut {NoStop}%
\bibitem [{\citenamefont {Turro}\ \emph {et~al.}(2022)\citenamefont {Turro}, \citenamefont {Roggero}, \citenamefont {Amitrano}, \citenamefont {Luchi}, \citenamefont {Wendt}, \citenamefont {Dubois}, \citenamefont {Quaglioni},\ and\ \citenamefont {Pederiva}}]{PhysRevA.105.022440}%
  \BibitemOpen
  \bibfield  {author} {\bibinfo {author} {\bibfnamefont {F.}~\bibnamefont {Turro}}, \bibinfo {author} {\bibfnamefont {A.}~\bibnamefont {Roggero}}, \bibinfo {author} {\bibfnamefont {V.}~\bibnamefont {Amitrano}}, \bibinfo {author} {\bibfnamefont {P.}~\bibnamefont {Luchi}}, \bibinfo {author} {\bibfnamefont {K.~A.}\ \bibnamefont {Wendt}}, \bibinfo {author} {\bibfnamefont {J.~L.}\ \bibnamefont {Dubois}}, \bibinfo {author} {\bibfnamefont {S.}~\bibnamefont {Quaglioni}},\ and\ \bibinfo {author} {\bibfnamefont {F.}~\bibnamefont {Pederiva}},\ }\bibfield  {title} {\bibinfo {title} {Imaginary-time propagation on a quantum chip},\ }\href {https://doi.org/10.1103/PhysRevA.105.022440} {\bibfield  {journal} {\bibinfo  {journal} {Phys. Rev. A}\ }\textbf {\bibinfo {volume} {105}},\ \bibinfo {pages} {022440} (\bibinfo {year} {2022})}\BibitemShut {NoStop}%
\bibitem [{\citenamefont {Jouzdani}\ \emph {et~al.}(2022)\citenamefont {Jouzdani}, \citenamefont {Johnson}, \citenamefont {Mucciolo},\ and\ \citenamefont {Stetcu}}]{PhysRevA.106.062435}%
  \BibitemOpen
  \bibfield  {author} {\bibinfo {author} {\bibfnamefont {P.}~\bibnamefont {Jouzdani}}, \bibinfo {author} {\bibfnamefont {C.~W.}\ \bibnamefont {Johnson}}, \bibinfo {author} {\bibfnamefont {E.~R.}\ \bibnamefont {Mucciolo}},\ and\ \bibinfo {author} {\bibfnamefont {I.}~\bibnamefont {Stetcu}},\ }\bibfield  {title} {\bibinfo {title} {Alternative approach to quantum imaginary time evolution},\ }\href {https://doi.org/10.1103/PhysRevA.106.062435} {\bibfield  {journal} {\bibinfo  {journal} {Phys. Rev. A}\ }\textbf {\bibinfo {volume} {106}},\ \bibinfo {pages} {062435} (\bibinfo {year} {2022})}\BibitemShut {NoStop}%
\bibitem [{\citenamefont {Preskill}(2018)}]{Preskill_2018}%
  \BibitemOpen
  \bibfield  {author} {\bibinfo {author} {\bibfnamefont {J.}~\bibnamefont {Preskill}},\ }\bibfield  {title} {\bibinfo {title} {Quantum computing in the {NISQ} era and beyond},\ }\href {https://doi.org/10.22331/q-2018-08-06-79} {\bibfield  {journal} {\bibinfo  {journal} {Quantum}\ }\textbf {\bibinfo {volume} {2}},\ \bibinfo {pages} {79} (\bibinfo {year} {2018})}\BibitemShut {NoStop}%
\bibitem [{\citenamefont {Peruzzo}\ \emph {et~al.}(2014)\citenamefont {Peruzzo}, \citenamefont {McClean}, \citenamefont {Shadbolt}, \citenamefont {Yung}, \citenamefont {Zhou}, \citenamefont {Love}, \citenamefont {Aspuru-Guzik},\ and\ \citenamefont {O'Brien}}]{Peruzzo2014}%
  \BibitemOpen
  \bibfield  {author} {\bibinfo {author} {\bibfnamefont {A.}~\bibnamefont {Peruzzo}}, \bibinfo {author} {\bibfnamefont {J.}~\bibnamefont {McClean}}, \bibinfo {author} {\bibfnamefont {P.}~\bibnamefont {Shadbolt}}, \bibinfo {author} {\bibfnamefont {M.-H.}\ \bibnamefont {Yung}}, \bibinfo {author} {\bibfnamefont {X.-Q.}\ \bibnamefont {Zhou}}, \bibinfo {author} {\bibfnamefont {P.~J.}\ \bibnamefont {Love}}, \bibinfo {author} {\bibfnamefont {A.}~\bibnamefont {Aspuru-Guzik}},\ and\ \bibinfo {author} {\bibfnamefont {J.~L.}\ \bibnamefont {O'Brien}},\ }\bibfield  {title} {\bibinfo {title} {A variational eigenvalue solver on a photonic quantum processor},\ }\href {https://doi.org/10.1038/ncomms5213} {\bibfield  {journal} {\bibinfo  {journal} {Nature Communications}\ }\textbf {\bibinfo {volume} {5}},\ \bibinfo {pages} {4213} (\bibinfo {year} {2014})}\BibitemShut {NoStop}%
\bibitem [{\citenamefont {McClean}\ \emph {et~al.}(2016)\citenamefont {McClean}, \citenamefont {Romero}, \citenamefont {Babbush},\ and\ \citenamefont {Aspuru-Guzik}}]{McClean_2016}%
  \BibitemOpen
  \bibfield  {author} {\bibinfo {author} {\bibfnamefont {J.~R.}\ \bibnamefont {McClean}}, \bibinfo {author} {\bibfnamefont {J.}~\bibnamefont {Romero}}, \bibinfo {author} {\bibfnamefont {R.}~\bibnamefont {Babbush}},\ and\ \bibinfo {author} {\bibfnamefont {A.}~\bibnamefont {Aspuru-Guzik}},\ }\bibfield  {title} {\bibinfo {title} {The theory of variational hybrid quantum-classical algorithms},\ }\href {https://doi.org/10.1088/1367-2630/18/2/023023} {\bibfield  {journal} {\bibinfo  {journal} {New Journal of Physics}\ }\textbf {\bibinfo {volume} {18}},\ \bibinfo {pages} {023023} (\bibinfo {year} {2016})}\BibitemShut {NoStop}%
\bibitem [{\citenamefont {Tilly}\ \emph {et~al.}(2022)\citenamefont {Tilly}, \citenamefont {Chen}, \citenamefont {Cao}, \citenamefont {Picozzi}, \citenamefont {Setia}, \citenamefont {Li}, \citenamefont {Grant}, \citenamefont {Wossnig}, \citenamefont {Rungger}, \citenamefont {Booth},\ and\ \citenamefont {Tennyson}}]{tilly2022variational}%
  \BibitemOpen
  \bibfield  {author} {\bibinfo {author} {\bibfnamefont {J.}~\bibnamefont {Tilly}}, \bibinfo {author} {\bibfnamefont {H.}~\bibnamefont {Chen}}, \bibinfo {author} {\bibfnamefont {S.}~\bibnamefont {Cao}}, \bibinfo {author} {\bibfnamefont {D.}~\bibnamefont {Picozzi}}, \bibinfo {author} {\bibfnamefont {K.}~\bibnamefont {Setia}}, \bibinfo {author} {\bibfnamefont {Y.}~\bibnamefont {Li}}, \bibinfo {author} {\bibfnamefont {E.}~\bibnamefont {Grant}}, \bibinfo {author} {\bibfnamefont {L.}~\bibnamefont {Wossnig}}, \bibinfo {author} {\bibfnamefont {I.}~\bibnamefont {Rungger}}, \bibinfo {author} {\bibfnamefont {G.~H.}\ \bibnamefont {Booth}},\ and\ \bibinfo {author} {\bibfnamefont {J.}~\bibnamefont {Tennyson}},\ }\href@noop {} {\bibinfo {title} {The variational quantum eigensolver: a review of methods and best practices}} (\bibinfo {year} {2022})\BibitemShut {NoStop}%
\bibitem [{\citenamefont {Lipkin}\ \emph {et~al.}(1965)\citenamefont {Lipkin}, \citenamefont {Meshkov},\ and\ \citenamefont {Glick}}]{LIPKIN1965188}%
  \BibitemOpen
  \bibfield  {author} {\bibinfo {author} {\bibfnamefont {H.}~\bibnamefont {Lipkin}}, \bibinfo {author} {\bibfnamefont {N.}~\bibnamefont {Meshkov}},\ and\ \bibinfo {author} {\bibfnamefont {A.}~\bibnamefont {Glick}},\ }\bibfield  {title} {\bibinfo {title} {Validity of many-body approximation methods for a solvable model: (i). exact solutions and perturbation theory},\ }\href {https://doi.org/https://doi.org/10.1016/0029-5582(65)90862-X} {\bibfield  {journal} {\bibinfo  {journal} {Nuclear Physics}\ }\textbf {\bibinfo {volume} {62}},\ \bibinfo {pages} {188} (\bibinfo {year} {1965})}\BibitemShut {NoStop}%
\bibitem [{\citenamefont {Agassi}\ \emph {et~al.}(1966)\citenamefont {Agassi}, \citenamefont {Lipkin},\ and\ \citenamefont {Meshkov}}]{agassi_validity_1966}%
  \BibitemOpen
  \bibfield  {author} {\bibinfo {author} {\bibfnamefont {D.}~\bibnamefont {Agassi}}, \bibinfo {author} {\bibfnamefont {H.~J.}\ \bibnamefont {Lipkin}},\ and\ \bibinfo {author} {\bibfnamefont {N.}~\bibnamefont {Meshkov}},\ }\bibfield  {title} {\bibinfo {title} {Validity of many-body approximation methods for a solvable model: ({IV}). {The} deformed {Hartree}-{Fock} solution},\ }\href {https://doi.org/10.1016/0029-5582(66)90540-2} {\bibfield  {journal} {\bibinfo  {journal} {Nuclear Physics}\ }\textbf {\bibinfo {volume} {86}},\ \bibinfo {pages} {321} (\bibinfo {year} {1966})}\BibitemShut {NoStop}%
\bibitem [{\citenamefont {Jordan}\ and\ \citenamefont {Wigner}(1928)}]{Jordan1928}%
  \BibitemOpen
  \bibfield  {author} {\bibinfo {author} {\bibfnamefont {P.}~\bibnamefont {Jordan}}\ and\ \bibinfo {author} {\bibfnamefont {E.}~\bibnamefont {Wigner}},\ }\bibfield  {title} {\bibinfo {title} {{\"U}ber das {P}aulische {{\"A}}quivalenzverbot},\ }\href {https://doi.org/10.1007/BF01331938} {\bibfield  {journal} {\bibinfo  {journal} {Zeitschrift f{\"u}r Physik}\ }\textbf {\bibinfo {volume} {47}},\ \bibinfo {pages} {631} (\bibinfo {year} {1928})}\BibitemShut {NoStop}%
\bibitem [{\citenamefont {Bravyi}\ and\ \citenamefont {Kitaev}(2002)}]{BRAVYI2002210}%
  \BibitemOpen
  \bibfield  {author} {\bibinfo {author} {\bibfnamefont {S.~B.}\ \bibnamefont {Bravyi}}\ and\ \bibinfo {author} {\bibfnamefont {A.~Y.}\ \bibnamefont {Kitaev}},\ }\bibfield  {title} {\bibinfo {title} {Fermionic quantum computation},\ }\href {https://doi.org/https://doi.org/10.1006/aphy.2002.6254} {\bibfield  {journal} {\bibinfo  {journal} {Annals of Physics}\ }\textbf {\bibinfo {volume} {298}},\ \bibinfo {pages} {210} (\bibinfo {year} {2002})}\BibitemShut {NoStop}%
\bibitem [{\citenamefont {Seeley}\ \emph {et~al.}(2012)\citenamefont {Seeley}, \citenamefont {Richard},\ and\ \citenamefont {Love}}]{Bravyi}%
  \BibitemOpen
  \bibfield  {author} {\bibinfo {author} {\bibfnamefont {J.~T.}\ \bibnamefont {Seeley}}, \bibinfo {author} {\bibfnamefont {M.~J.}\ \bibnamefont {Richard}},\ and\ \bibinfo {author} {\bibfnamefont {P.~J.}\ \bibnamefont {Love}},\ }\bibfield  {title} {\bibinfo {title} {{The Bravyi-Kitaev transformation for quantum computation of electronic structure}},\ }\href {https://doi.org/10.1063/1.4768229} {\bibfield  {journal} {\bibinfo  {journal} {The Journal of Chemical Physics}\ }\textbf {\bibinfo {volume} {137}},\ \bibinfo {pages} {224109} (\bibinfo {year} {2012})}\BibitemShut {NoStop}%
\bibitem [{\citenamefont {Pesce}\ and\ \citenamefont {Stevenson}(2021)}]{pesce_h2zixy_2021}%
  \BibitemOpen
  \bibfield  {author} {\bibinfo {author} {\bibfnamefont {R.~M.~N.}\ \bibnamefont {Pesce}}\ and\ \bibinfo {author} {\bibfnamefont {P.~D.}\ \bibnamefont {Stevenson}},\ }\href@noop {} {\bibinfo {title} {{H2ZIXY}: {Pauli} spin matrix decomposition of real symmetric matrices}} (\bibinfo {year} {2021}),\ \Eprint {https://arxiv.org/abs/2111.00627} {arXiv:2111.00627} \BibitemShut {NoStop}%
\bibitem [{\citenamefont {Bergholm}\ \emph {et~al.}(2022)\citenamefont {Bergholm}, \citenamefont {Izaac}, \citenamefont {Schuld}, \citenamefont {Gogolin}, \citenamefont {Ahmed}, \citenamefont {Ajith}, \citenamefont {Alam}, \citenamefont {Alonso-Linaje}, \citenamefont {AkashNarayanan}, \citenamefont {Asadi}, \citenamefont {Arrazola}, \citenamefont {Azad}, \citenamefont {Banning}, \citenamefont {Blank}, \citenamefont {Bromley}, \citenamefont {Cordier}, \citenamefont {Ceroni}, \citenamefont {Delgado}, \citenamefont {Matteo}, \citenamefont {Dusko}, \citenamefont {Garg}, \citenamefont {Guala}, \citenamefont {Hayes}, \citenamefont {Hill}, \citenamefont {Ijaz}, \citenamefont {Isacsson}, \citenamefont {Ittah}, \citenamefont {Jahangiri}, \citenamefont {Jain}, \citenamefont {Jiang}, \citenamefont {Khandelwal}, \citenamefont {Kottmann}, \citenamefont {Lang}, \citenamefont {Lee}, \citenamefont {Loke}, \citenamefont {Lowe}, \citenamefont {McKiernan}, \citenamefont {Meyer}, \citenamefont {Montañez-Barrera}, \citenamefont
  {Moyard}, \citenamefont {Niu}, \citenamefont {O'Riordan}, \citenamefont {Oud}, \citenamefont {Panigrahi}, \citenamefont {Park}, \citenamefont {Polatajko}, \citenamefont {Quesada}, \citenamefont {Roberts}, \citenamefont {Sá}, \citenamefont {Schoch}, \citenamefont {Shi}, \citenamefont {Shu}, \citenamefont {Sim}, \citenamefont {Singh}, \citenamefont {Strandberg}, \citenamefont {Soni}, \citenamefont {Száva}, \citenamefont {Thabet}, \citenamefont {Vargas-Hernández}, \citenamefont {Vincent}, \citenamefont {Vitucci}, \citenamefont {Weber}, \citenamefont {Wierichs}, \citenamefont {Wiersema}, \citenamefont {Willmann}, \citenamefont {Wong}, \citenamefont {Zhang},\ and\ \citenamefont {Killoran}}]{bergholm2022pennylane}%
  \BibitemOpen
  \bibfield  {author} {\bibinfo {author} {\bibfnamefont {V.}~\bibnamefont {Bergholm}}, \bibinfo {author} {\bibfnamefont {J.}~\bibnamefont {Izaac}}, \bibinfo {author} {\bibfnamefont {M.}~\bibnamefont {Schuld}}, \bibinfo {author} {\bibfnamefont {C.}~\bibnamefont {Gogolin}}, \bibinfo {author} {\bibfnamefont {S.}~\bibnamefont {Ahmed}}, \bibinfo {author} {\bibfnamefont {V.}~\bibnamefont {Ajith}}, \bibinfo {author} {\bibfnamefont {M.~S.}\ \bibnamefont {Alam}}, \bibinfo {author} {\bibfnamefont {G.}~\bibnamefont {Alonso-Linaje}}, \bibinfo {author} {\bibfnamefont {B.}~\bibnamefont {AkashNarayanan}}, \bibinfo {author} {\bibfnamefont {A.}~\bibnamefont {Asadi}}, \bibinfo {author} {\bibfnamefont {J.~M.}\ \bibnamefont {Arrazola}}, \bibinfo {author} {\bibfnamefont {U.}~\bibnamefont {Azad}}, \bibinfo {author} {\bibfnamefont {S.}~\bibnamefont {Banning}}, \bibinfo {author} {\bibfnamefont {C.}~\bibnamefont {Blank}}, \bibinfo {author} {\bibfnamefont {T.~R.}\ \bibnamefont {Bromley}}, \bibinfo {author} {\bibfnamefont {B.~A.}\
  \bibnamefont {Cordier}}, \bibinfo {author} {\bibfnamefont {J.}~\bibnamefont {Ceroni}}, \bibinfo {author} {\bibfnamefont {A.}~\bibnamefont {Delgado}}, \bibinfo {author} {\bibfnamefont {O.~D.}\ \bibnamefont {Matteo}}, \bibinfo {author} {\bibfnamefont {A.}~\bibnamefont {Dusko}}, \bibinfo {author} {\bibfnamefont {T.}~\bibnamefont {Garg}}, \bibinfo {author} {\bibfnamefont {D.}~\bibnamefont {Guala}}, \bibinfo {author} {\bibfnamefont {A.}~\bibnamefont {Hayes}}, \bibinfo {author} {\bibfnamefont {R.}~\bibnamefont {Hill}}, \bibinfo {author} {\bibfnamefont {A.}~\bibnamefont {Ijaz}}, \bibinfo {author} {\bibfnamefont {T.}~\bibnamefont {Isacsson}}, \bibinfo {author} {\bibfnamefont {D.}~\bibnamefont {Ittah}}, \bibinfo {author} {\bibfnamefont {S.}~\bibnamefont {Jahangiri}}, \bibinfo {author} {\bibfnamefont {P.}~\bibnamefont {Jain}}, \bibinfo {author} {\bibfnamefont {E.}~\bibnamefont {Jiang}}, \bibinfo {author} {\bibfnamefont {A.}~\bibnamefont {Khandelwal}}, \bibinfo {author} {\bibfnamefont {K.}~\bibnamefont {Kottmann}},
  \bibinfo {author} {\bibfnamefont {R.~A.}\ \bibnamefont {Lang}}, \bibinfo {author} {\bibfnamefont {C.}~\bibnamefont {Lee}}, \bibinfo {author} {\bibfnamefont {T.}~\bibnamefont {Loke}}, \bibinfo {author} {\bibfnamefont {A.}~\bibnamefont {Lowe}}, \bibinfo {author} {\bibfnamefont {K.}~\bibnamefont {McKiernan}}, \bibinfo {author} {\bibfnamefont {J.~J.}\ \bibnamefont {Meyer}}, \bibinfo {author} {\bibfnamefont {J.~A.}\ \bibnamefont {Montañez-Barrera}}, \bibinfo {author} {\bibfnamefont {R.}~\bibnamefont {Moyard}}, \bibinfo {author} {\bibfnamefont {Z.}~\bibnamefont {Niu}}, \bibinfo {author} {\bibfnamefont {L.~J.}\ \bibnamefont {O'Riordan}}, \bibinfo {author} {\bibfnamefont {S.}~\bibnamefont {Oud}}, \bibinfo {author} {\bibfnamefont {A.}~\bibnamefont {Panigrahi}}, \bibinfo {author} {\bibfnamefont {C.-Y.}\ \bibnamefont {Park}}, \bibinfo {author} {\bibfnamefont {D.}~\bibnamefont {Polatajko}}, \bibinfo {author} {\bibfnamefont {N.}~\bibnamefont {Quesada}}, \bibinfo {author} {\bibfnamefont {C.}~\bibnamefont {Roberts}},
  \bibinfo {author} {\bibfnamefont {N.}~\bibnamefont {Sá}}, \bibinfo {author} {\bibfnamefont {I.}~\bibnamefont {Schoch}}, \bibinfo {author} {\bibfnamefont {B.}~\bibnamefont {Shi}}, \bibinfo {author} {\bibfnamefont {S.}~\bibnamefont {Shu}}, \bibinfo {author} {\bibfnamefont {S.}~\bibnamefont {Sim}}, \bibinfo {author} {\bibfnamefont {A.}~\bibnamefont {Singh}}, \bibinfo {author} {\bibfnamefont {I.}~\bibnamefont {Strandberg}}, \bibinfo {author} {\bibfnamefont {J.}~\bibnamefont {Soni}}, \bibinfo {author} {\bibfnamefont {A.}~\bibnamefont {Száva}}, \bibinfo {author} {\bibfnamefont {S.}~\bibnamefont {Thabet}}, \bibinfo {author} {\bibfnamefont {R.~A.}\ \bibnamefont {Vargas-Hernández}}, \bibinfo {author} {\bibfnamefont {T.}~\bibnamefont {Vincent}}, \bibinfo {author} {\bibfnamefont {N.}~\bibnamefont {Vitucci}}, \bibinfo {author} {\bibfnamefont {M.}~\bibnamefont {Weber}}, \bibinfo {author} {\bibfnamefont {D.}~\bibnamefont {Wierichs}}, \bibinfo {author} {\bibfnamefont {R.}~\bibnamefont {Wiersema}}, \bibinfo {author}
  {\bibfnamefont {M.}~\bibnamefont {Willmann}}, \bibinfo {author} {\bibfnamefont {V.}~\bibnamefont {Wong}}, \bibinfo {author} {\bibfnamefont {S.}~\bibnamefont {Zhang}},\ and\ \bibinfo {author} {\bibfnamefont {N.}~\bibnamefont {Killoran}},\ }\href@noop {} {\bibinfo {title} {Pennylane: Automatic differentiation of hybrid quantum-classical computations}} (\bibinfo {year} {2022}),\ \Eprint {https://arxiv.org/abs/1811.04968} {arXiv:1811.04968} \BibitemShut {NoStop}%
\bibitem [{\citenamefont {Kandala}\ \emph {et~al.}(2017)\citenamefont {Kandala}, \citenamefont {Mezzacapo}, \citenamefont {Temme}, \citenamefont {Takita}, \citenamefont {Brink}, \citenamefont {Chow},\ and\ \citenamefont {Gambetta}}]{Kandala2017}%
  \BibitemOpen
  \bibfield  {author} {\bibinfo {author} {\bibfnamefont {A.}~\bibnamefont {Kandala}}, \bibinfo {author} {\bibfnamefont {A.}~\bibnamefont {Mezzacapo}}, \bibinfo {author} {\bibfnamefont {K.}~\bibnamefont {Temme}}, \bibinfo {author} {\bibfnamefont {M.}~\bibnamefont {Takita}}, \bibinfo {author} {\bibfnamefont {M.}~\bibnamefont {Brink}}, \bibinfo {author} {\bibfnamefont {J.~M.}\ \bibnamefont {Chow}},\ and\ \bibinfo {author} {\bibfnamefont {J.~M.}\ \bibnamefont {Gambetta}},\ }\bibfield  {title} {\bibinfo {title} {Hardware-efficient variational quantum eigensolver for small molecules and quantum magnets},\ }\href {https://doi.org/10.1038/nature23879} {\bibfield  {journal} {\bibinfo  {journal} {Nature}\ }\textbf {\bibinfo {volume} {549}},\ \bibinfo {pages} {242} (\bibinfo {year} {2017})}\BibitemShut {NoStop}%
\bibitem [{\citenamefont {Li}\ and\ \citenamefont {Benjamin}(2017)}]{PhysRevX.7.021050}%
  \BibitemOpen
  \bibfield  {author} {\bibinfo {author} {\bibfnamefont {Y.}~\bibnamefont {Li}}\ and\ \bibinfo {author} {\bibfnamefont {S.~C.}\ \bibnamefont {Benjamin}},\ }\bibfield  {title} {\bibinfo {title} {Efficient variational quantum simulator incorporating active error minimization},\ }\href {https://doi.org/10.1103/PhysRevX.7.021050} {\bibfield  {journal} {\bibinfo  {journal} {Phys. Rev. X}\ }\textbf {\bibinfo {volume} {7}},\ \bibinfo {pages} {021050} (\bibinfo {year} {2017})}\BibitemShut {NoStop}%
\end{thebibliography}%

\end{document}